\documentclass[]{aa}
\usepackage{natbib}
\usepackage{psfig}
\def\158{$\lambda 158\mu$}
\def\63{$\lambda 63\mu$}
\def\etc{{\it etc.}}

\def\EBV{E$_{B-V}$}
\def\HH{{\rm H}_2}
\def\nH2{{\rm n}({\rm H}_2)}
\def\NH2{{\rm N}({\rm H}_2)}
\def\pccc{{\rm cm}^{-3}} \def\pcc{{\rm cm}^{-2}}

\def\Tstar#1 {\mbox{${\rm T}_{\rm #1}^*$}}
\def\Tsub#1 {\mbox{${\rm T}_{\rm #1}$}}
\def\TK  {\Tsub K }

\def\Tsp {\Tsub sp }

\def\Tcmb{\Tsub cmb }

\def\arcmin{\mbox{$^{\prime}$}}

\def\p{$^+$}

\def\hcop{\mbox{{HCO\p}}}

\def\h13cop{\mbox{{H$^{13}$CO\p}}}

\def\c3h2{\mbox{C$_3$H$_2$}}
 
 \def\R0{R$_0$}

\def\ddeg{{}^\circ\kern-.1em}

\def\ps{\mbox{s$^{-1}$}}

\def\E#1{\,10^{#1}}
\def\P#1,{$\nH2\TK~=~#1\times~10^4~\pccc$~K}
\def\ec#1,#2,#3,{#1\,(#2)\E{#3}}

\def\H3{\mbox{H$_3$}}
\def\Lya{\mbox{Ly-$\alpha$}}

\def\DLAS{damped Lyman-$\alpha$ systems}
\title{Two-phase equilibrium and molecular
hydrogen formation in damped Lyman-alpha systems}

\titlerunning{$\HH$ formation in \DLAS}

\author{H. Liszt\inst{1}}
\institute{National Radio Astronomy Observatory,
           520 Edgemont Road,
           Charlottesville, VA,
           USA 22903-2475}

\begin{document}
\date{received \today}
\offprints{H. S. Liszt}
\mail{hliszt@nrao.edu}

\abstract{
Molecular hydrogen is quite underabundant in \DLAS\ at high 
redshift, when compared to the interstellar medium
near the Sun.  This has been interpreted as implying that the
gas in \DLAS\ is warm. like the nearby neutral intercloud medium, 
rather than cool, as in the clouds which give rise to most H I 
absorption in the Milky Way.  Other lines of evidence 
suggest that the gas in \DLAS\ -- in whole or part -- is actually 
cool;  spectroscopy of neutral and ionized carbon, discussed here, 
shows that the \DLAS\ observed at lower redshift z $<$ 2.3 are 
largely cool, while those seen at z $>$ 2.8 are warm (though not
devoid of $\HH$).  To interpret the observations 
of carbon and hydrogen  we constructed detailed numerical models of 
$\HH$ formation under the conditions of two-phase thermal equilibrium, 
like those which account for conditions near the Sun, but with varying 
metallicity, dust-gas ratio, $etc$.  We find that the 
low metallicity of \DLAS\ is enough to suppress $\HH$ formation by many 
orders of magnitude even in cool diffuse clouds, as long as the ambient 
optical/uv radiation field is not too small.  For very low metallicity 
and under the most diffuse conditions, $\HH$ formation will  
be dominated by slow gas-phase processes not involving grains, and a 
minimum molecular fraction in the range $10^{-8}-10^{-7}$ is expected.
}
\maketitle

\section {Introduction.}

Perhaps it is something of a paradox, but large ground-based
optical telescopes like Keck and VLT routinely do absorption-line 
spectroscopy of the gas in high-redshift objects which surpasses what 
can be done for the interstellar medium (ISM) of our own Galaxy near 
the Sun.  Viewing neutral gas at high redshift makes the dominant ion 
stages (C II, Fe II, \etc) accessible,  to say nothing of Lyman series 
in H I and D I and the Lyman and Werner bands of $\HH$.  The high spectral 
resolution and large collecting areas of modern ground-based 
instrumentation cannot easily be matched in space-based instruments, 
and certainly not for the same cost.

Thus we may have exquisitely detailed and sensitive spectra of systems
which cannot be imaged, whose nature is therefore left to be inferred 
from their patterns of gas kinematics 
\citep{WolPro97,WolPro98,HaeSte+98,LedPet+98,McDMir99} and ionization 
\citep{WolPro00a,WolPro00b}. So it is with most \DLAS, defined as those 
absorption-line systems having N(H I) $\ge 2\times 10^{20}~\pcc$ 
when seen against the emission of background QSO's. Although well-formed 
galaxian systems can and do harbor some of them at low redshift,
\DLAS\ are for the most part believed to be protogalactic objects, perhaps 
in disk systems (Prochaska \& Wolfe) or perhaps in the ongoing merger of 
protogalactic clumps \citep{HaeSte+98}.  The numbers are such that \DLAS\ 
seem to contain at least as many baryons as can be found in the local Universe 
now, and it is of great interest to understand how these baryons become 
recognizable nearby objects within a relatively short redshift interval.

The absorption-line gas in \DLAS\ is easily discussed in the same terms, 
using the  same physical processes, that are employed locally.  
Evidence that the rest-frame optical/uv  radiation field is comparable
to that near the Sun \citep{LevDes+02,MolLev+02,PetSri+00} and the presence 
of metals at a low but non-negligible level 0.1 - 0.01 Solar, makes it 
possible (or, perhaps, merely hopeful?) to talk about the 
``interstellar medium'' in these systems.  The low metallicity and general 
underabundance of dust (which is not directly observed in any one 
object but can be inferred statistically \citep{FalPei93,PeiFal+91} or 
from patterns of gaseous abundances \citep{BoiLeB+98}) may render the 
gaseous medium in \DLAS\ only more extreme versions of those in local 
dwarf systems like the LMC and SMC \footnote{\cite{BoiLeB+98} show that 
there there is a bias against finding high metallicity \DLAS\ at high 
N(H I), because the background sources are extinguished.} .
But it has also been argued that the apparent recognizability 
of patterns in the absorption spectra has been over-interpreted
\citep{IzoSch+01}, and 
that the similarity of intermediate and low ion kinematics (Al III and 
C II or Fe II; C IV and C II behave differently) could mean that
substantial ionization corrections to the metallicity are needed
(however, see \cite{VlaCen+01} for a contrary opinion).

At the present time there seem to be several lines of evidence suggesting 
that, unlike the local ISM (where low-altitude neutral gas is perhaps
2/3 cool and 1/3 warm and the overall ratio including high-altitude
material is 1/2 and 1/2)  the gas in \DLAS\ is more predominantly warm. 
Examples include the high H I spin temperatures inferred from comparison 
of $\lambda21$cm and Lyman-$\alpha$ absorption
\citep{WolDav79,CarLan+96a,CarLan+96b,CheKan00,KanChe01}
(but see \cite{LanBri+00}), the similarity of low and intermediate ion 
kinematics mentioned in the preceding paragraph, and the very low column 
densities of $\HH$ discussed here.  Along lines of sight with reddening  
\EBV $>$ 0.05 mag (N(H) $> 3 \times 10^{20}~\pcc$) in local $Copernicus$ 
spectra, it is always the case that a few percent or more of the neutral 
hydrogen is molecular.  By contrast, the molecular hydrogen fraction in 
\DLAS\ is 2-4 orders of magnitude lower ($i.e. 10^{-4} - 10^{-6})$, which can 
be interpreted as meaning that the gas temperature must be above 3000 K 
\citep{PetSri+00} \cite{LanWol+89} used N($\HH$)/N(H) to constrain 
the dust/gas ratio toward Q1337+113, similar to the approach taken
in this work.

Here we consider the formation of molecular hydrogen in a gas which is in 
two-phase thermal equilibrium at low metallicity.  Perhaps because $\HH$ 
has been observable so rarely in the local ISM there is not a big literature 
on this subject, but the extant 1970's-era $Copernicus$ observations
are well-explained in this way \citep{LisLuc00} using modern shielding 
factors for radiative dissociation \citep{LeeHer+96}.  The only surprise 
(if there indeed is one) is the low densities that are required to start 
$\HH$ formation locally, and the fact that even a ``standard'' H I cloud 
\citep{Spi78} should have a molecular fraction of 10-30\% deep inside.  
Local diffuse clouds also have surprisingly high abundances of complex 
polyatomic species which follow immediately upon the presence of $\HH$ 
\citep{LisLuc96,LucLis96}, a phenomenon which is not well understood but 
which can be used to account for the observed abundancess of simpler 
species such as CO \citep{LisLuc00}.

\begin{figure*}
\psfig{figure=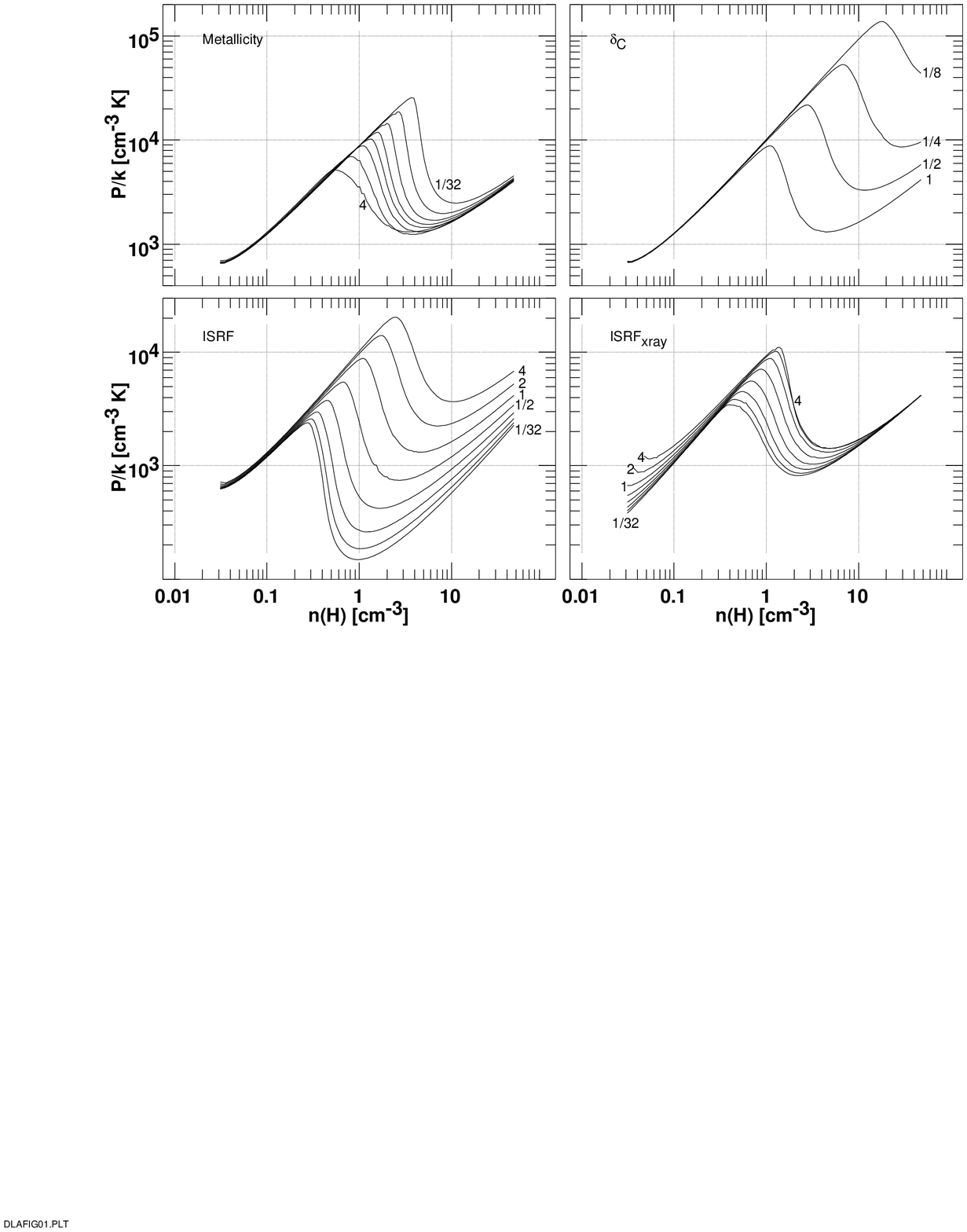,height=12.7cm}
\caption[]{Ionization and thermal equilibrium calculations in atomic gas.
The pressure P/k is shown as a function of the density of H-nuclei n(H), 
for four sets of parametric variations. a) Upper left; the metallicity,
(dust/gas, C/H, O/H $etc.$) varies in steps of 2 from 4 times to 1/32 times 
its reference value;  b)  Upper right, carbon and oxygen are removed 
(depleted) from the gas phase in steps of 2; c) Lower left, the 
``interstellar'' (ambient) radiation field (ISRF) varies from 4 times to 
1/32 times the reference value; d) Lower right, the impinging flux of soft 
X-rays is scaled.}
\end{figure*}

In the course of this work, we consider the ionization and 
fine-structure excitation of carbon, and the inferences which may be
drawn from observations of carbon in the \DLAS.  While the importance 
of carbon to the physical state of the gas and the conditions for 
forming $\HH$ cannot be stressed too highly, the extended discussion
here in fact arose because of something of a coincidence.  Searching the 
literature, it quickly became evident that there is a high degree of 
overlap between the two relatively scant datasets for carbon (chiefly,
C I, C II, and C II*) and $\HH$.

The plan of the current discussion is as follows.  In Sect. 2 we lay
out the basics of a calculation of two-phase equilibrium essentially
following \cite{WolHol+95}.  We examine the sensitivity of two-phase
equilibrium to variations in abundance, depletion, incident radiation
and the like, in order to extract from the calculations those aspects 
which can be related to existing absorption line data on \DLAS. 
In Sect. 3 we compare the results of these calculations with observed
hydrogen and carbon column densities: we show that the observations 
are consistent with an origin in largely cool gas for the \DLAS\ at lower
z (z $<$ 2.3) and in warm gas at higher redshift. In Sect. 
4 we describe calculations of the $\HH$ abundance in both warm and
cool diffuse gas, under conditions of varying metallicty, $etc.$, employing 
(slow) gas-phase processes to explore the minimum expected amounts of 
molecular gas, and the more usual grain-catalysis (as in \cite{LisLuc00}) 
for cooler regions of higher molecular fraction.  We also compute the 
variation of the molecular hydrogen fraction in small gas clots of constant 
density. From this, it follows that the low metallicities of \DLAS\ are by
themselves sufficient to cause decreases in the molecular fraction
by many orders of magnitude, even if cool neutral clouds are present.

\section{Two-phase equilibrium of the ISM}

\subsection{Particulars and parameter sensitivities}

The ISM is so complicated that any model of it is bound to be heavily 
idealized and stylized, especially the assumption of a strict equilibrium.  
But some aspects, especially the presence of phases -- discrete regimes of 
density and ionization -- seem robust to variations in the underlying 
parameters.  The existence of multiphase gas seems to be quite general, 
highly conserved across space and time.

The basis of the present work is a calculation of two-phase equilibrium 
following work on the local ISM by \cite{WolHol+95}.  Notable constraints on 
the model locally are the thermal pressure range P/k = $10^3 - 10^4~\pccc$ K 
observed ($via$ C I and C I*) in neutral gas locally by \cite{JenJur+83}, 
and the electron density in warm gas at the Solar radius 
n$_{\rm e} \approx 0.02~\pccc$ \citep{TayCor93}.  As noted by \cite{WolHol+95}, 
strict two-phase equilibrium 
is an idealized goal toward which the ISM may tend, but true equilibrium
is hard to attain, especially in warm gas where the cooling timescales
are long.

\begin{figure*}
\psfig{figure=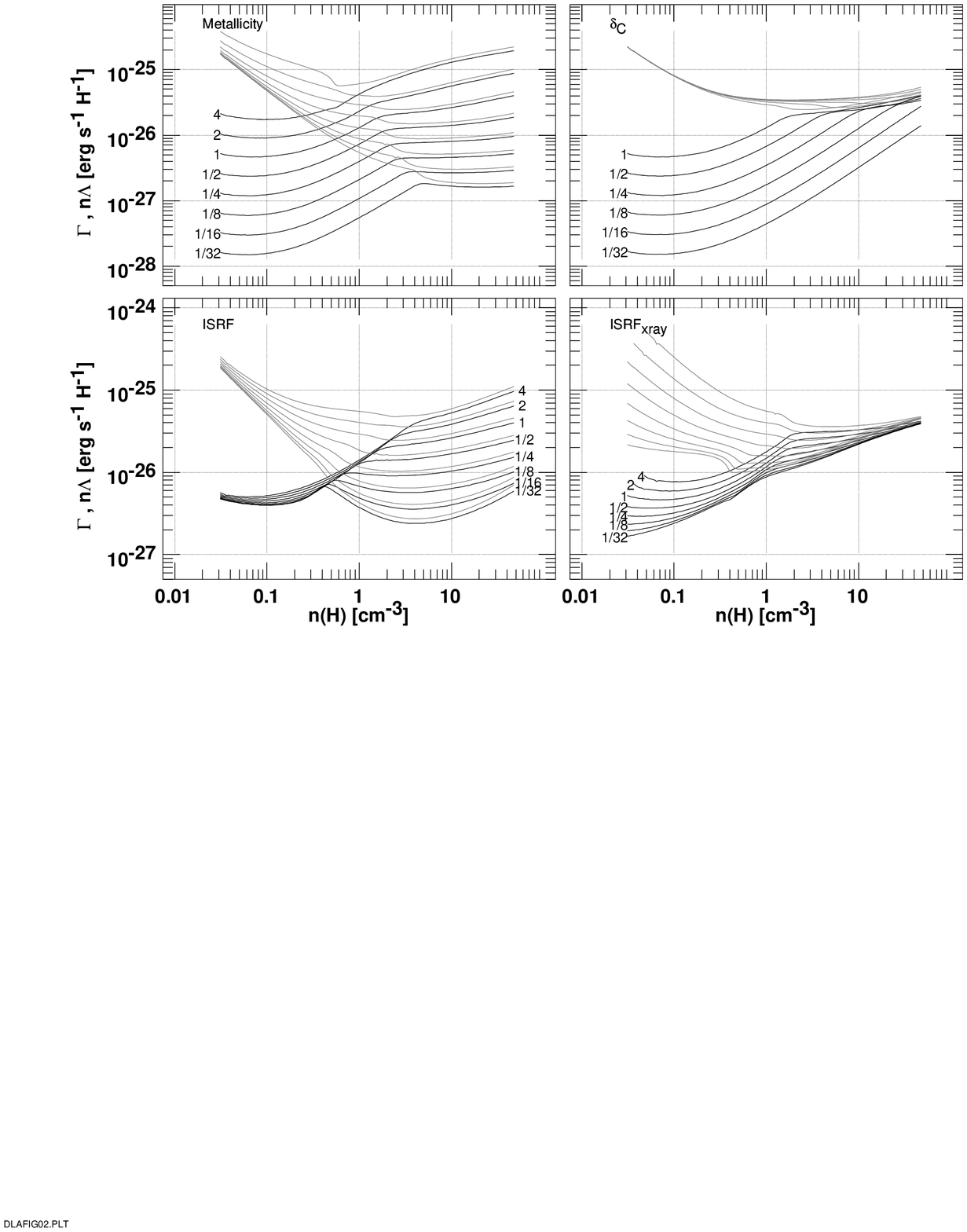,height=12.7cm}
\caption[]{As in Fig. 1, but the quantities shown in each panel are
the energy loss rate n$\Lambda$ due to radiation in the C II* \158\ cooling 
transition (lower, full curves) $\Gamma$, the total energy input to the
gas (upper, shaded curves), both in units of erg \ps\ H$^{-1}$.  In cool
gas (to the right), C II is by far the dominant coolant; in warm gas, most of 
the cooling occurs in the Ly-$\alpha$ line.  The mean C II cooling rate for 
the Milky Way is $2.65 \times 10^{-26}$ erg \ps\ H$^{-1}$ \citep{BenFix+94}}.
\end{figure*}

Basically, this "standard" model of the local diffuse ISM is driven by 
the heating due (in increasing order of importance in warm gas) to 
cosmic-rays, soft X-rays, and -- the dominant mechanism in both warm and 
cool gas -- the photoelectric effect on small grains \citep{BakTie94}.  
The charge state of the small grains, their heating rates and the 
recombination of ions on them (which dominates over gas-phase processes 
at moderate densities in diffuse gas) are all very sensitive to the 
electron density, so local thermal and ionization equilbria are 
tightly coupled.  Given that carbon is the main source of cooling
in cooler gas (O I to a lesser extent) and an important source of free 
electrons, it follows that the equilibrium conditions and the 
condition of carbon in the gas are very tightly coupled.

We recently used this code \citep{Lis01} to discuss the
behaviour of the H I spin temperature in local warm gas (the H I there
is not generally thermalized by collisions in multiphase equilbrium)
and to calculate the abundances of $\HH$ and CO in denser diffuse gas
\citep{LisLuc00}.  The latter reference describes the model in more
detail than will be given here, especially the various collisional
processes which drive fine-structure cooling in the gas (they are,
of course also discussed in the original reference by \cite{WolHol+95}).

Our basic calculation differs from that of \cite{WolHol+95} in two minor
ways which are subsumed by the extent of our parameter variations.
The earlier authors took the locally-determined soft X-ray flux from 
\cite{GarNou+92} and decomposed it into three components, each of which
they represented by a (physically-motivated) plasma emissivity
which was then separately attenuated (or not) by an assumed column of warm
gas; the three contributions were then summed to provide the assumed
incident X-ray spectrum at a typical location.  Their final spectrum 
is characterized by a quantity they called Nw, the overall column of
warm gas attenuating part of the incident spectrum
\footnote{The soft X-ray flux is so fragile that any cool gas interposed 
would absorb all of it, given that cool gas does not appear in 
small increments} 
whose standard value they took to be Nw $= 10^{19}$ H atoms $\pcc.$  We 
used a simpler representation of the soft X-ray flux, whereby the 
observed spectrum is attenuated by a column of neutral gas Nw directly.
Our standard value of  Nw $= 2 \times 10^{19}$ H atoms $\pcc$
produces a pressure-density curve which differs little from that
of \cite{WolHol+95}.

We also used a somewhat different set of reference (Solar) atomic 
abundances whereby log(C/H) = -3.44, log(O/C) = 0.32, namely those 
which accompanied the distribution of the soft X-ray absorption cross-
sections from \citep{BalMcC+92}, which both we and \cite{WolHol+95} used
(in their most recently updated version).  In both calculations the 
reference model has (following \cite{WolHol+95}) little or no
depletion of the gas phase oxygen and carbon onto grains, which 
might be acceptable for warm gas (where carbon does not bear the 
brunt of the cooling or contribute many of the electrons, see below) 
but is unlikely to be reasonable in local cool gas; if there are grains, 
they have to be made of something!  The gas phase depletion is considered 
an adjustable parameter here, independent of the metallicity.  

The net effect of these two differences is slight.  With the exception of
one system, the observations show that carbon is present in the
gas in the same proportion as other species which are typically
undepleted locally, $i.e.$ there is low metallicity, but little
global carbon depletion.

\begin{figure*}
\psfig{figure=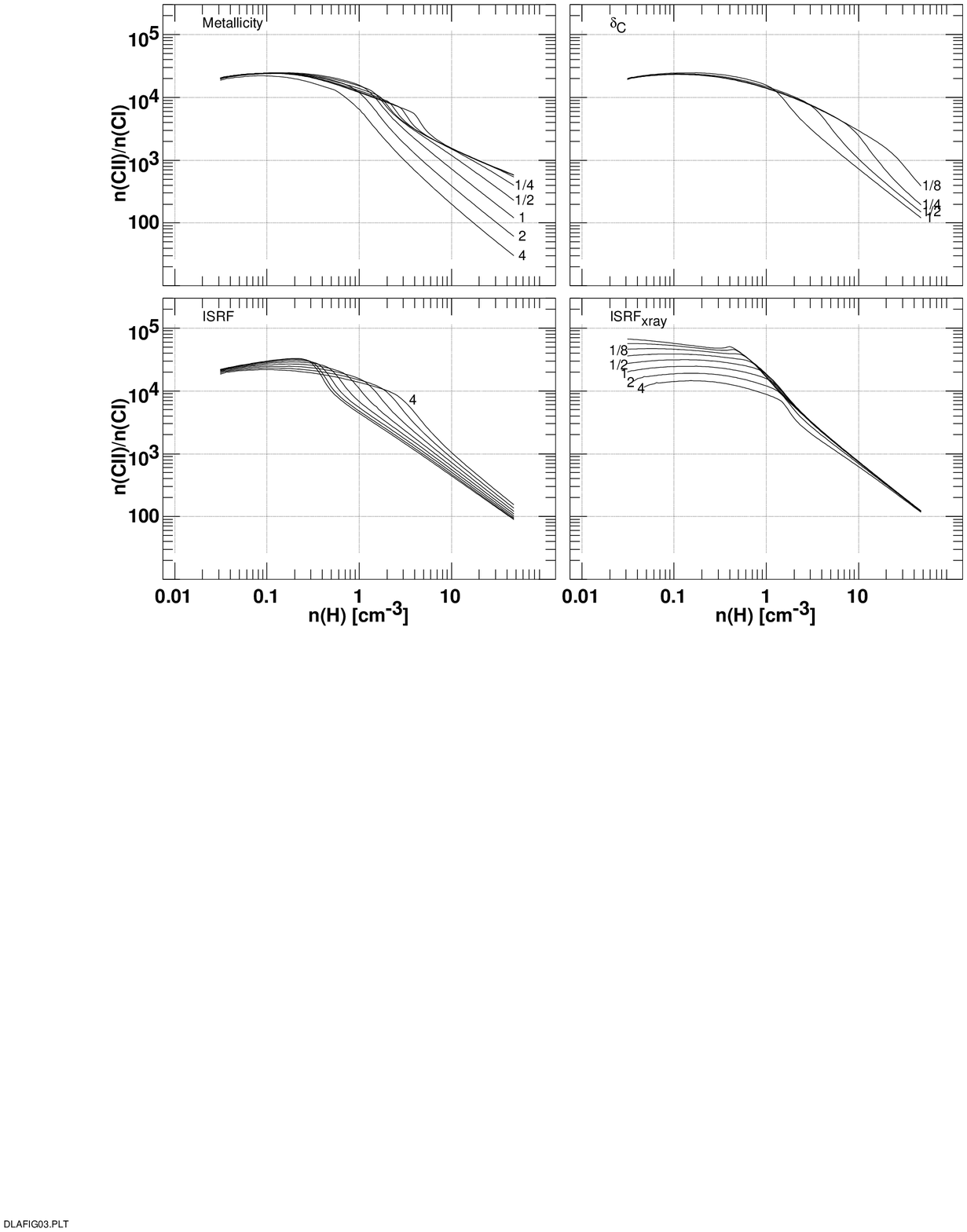,height=12.7cm}
\caption[]{As in Fig. 1 and Fig 2, but the quantity shown in each
panel here is the ratio of once-ionized to neutral carbon 
{n(C II)/n(C I)}.}
\end{figure*}

%
%

In Fig. 1  we show the basics of two-phase equilibrium and its 
sensitivity to some assumed parameter variations.  In these curves
of pressure $vs.$ density (one assumes a density, calculates the
ionization and thermal equilibrium, sums the particle densities
and multiplies by the derived temperature), the regions where
dP/dn $< 0$ are held to be unstable and unlikely to occur in nature. 
Those pressures for which there are two densities n\arcmin\ having 
dP/dn$|_{n'} > 0$ are those for which ``two-phase'' equilibrium
is possible.   Two-phase heating and cooling calculations by themselves
furnish only the possibility of multi-phase equilibrium; \cite{Hen00} and 
\cite{KriNor02} elaborate on some processes by which phase transitions and 
multi-phase equilibrium are actually brought about.

The curve labelled ``1'' in all panels is the same (reference) 
model; perhaps most clearly at upper right it is apparent that the 
reference model provides for two-phase equilibrium over precisely the 
pressure range which is observed in local gas \citep{JenJur+83}. Typically 
(but see below and the lower-left panel), only warm (7000 K - 10000 K) 
neutral gas appears if the density and pressure are below the two-phase 
regime and only cold (below 1000 K) neutral gas appears if the pressure 
and density are larger.  If conditions are such that the warm gas may 
persist up to higher density and pressure, or when the cold medium cannot 
exist except at higher density and pressure, the medium is more likely 
to be warm.

\begin{figure}
\psfig{figure=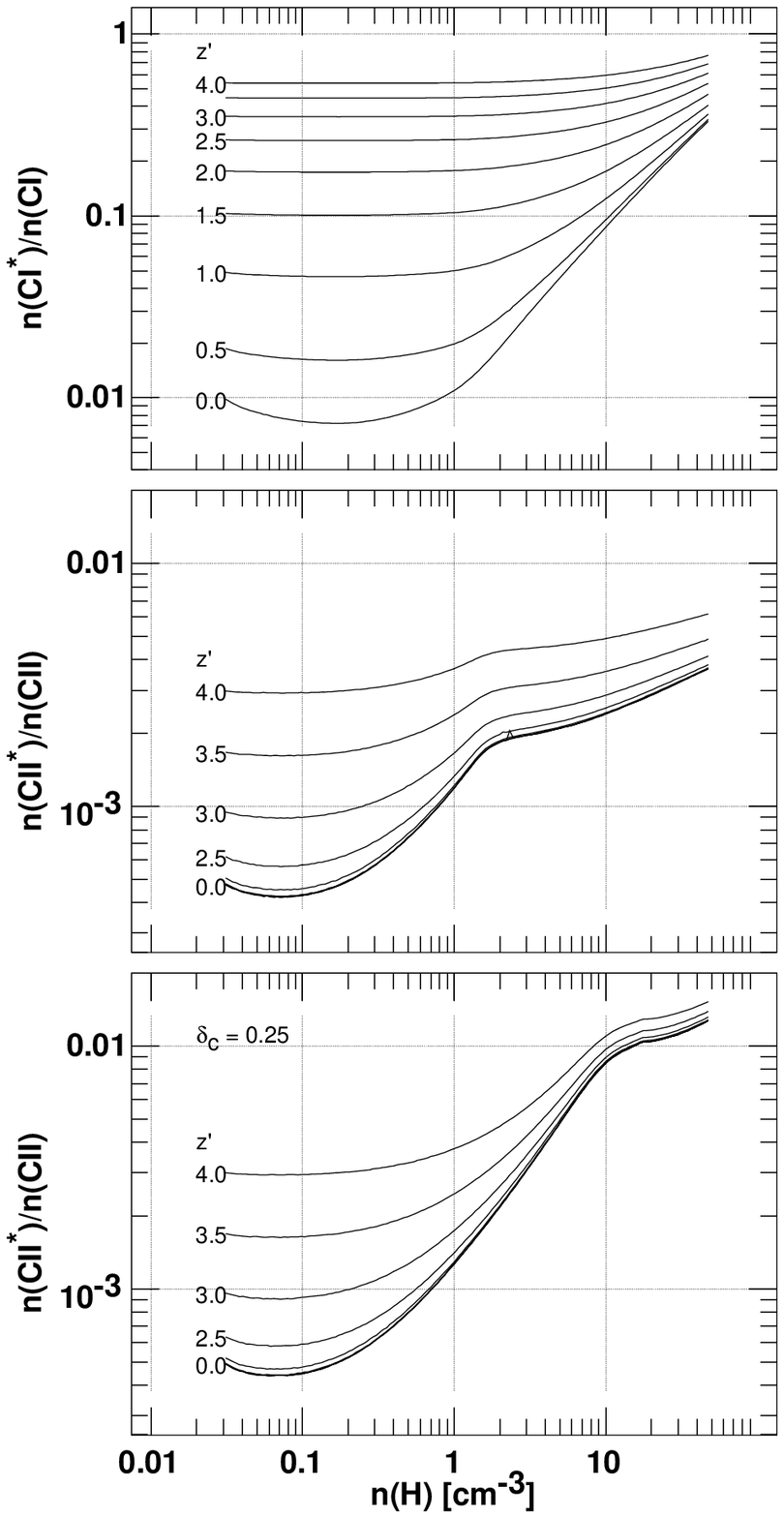,height=17cm}
\caption[]{Fine-structure excitation calculations for varying
\Tcmb = z\arcmin$\times$2.73 K.  Top: the ratio of first-excited to
ground-state populations in neutral carbon, CI.  Middle, the ratio
of excited to ground state populations in once-ionized carbon, C II.
Bottom, as in the middle panel but for the case where carbon is
depleted by a factor 4 from the gas-phase, relative to the standard
model.}
\end{figure}

At the top left, the metallicity is varied by factors of 4 to 1/32 
relative to the reference value ; these may be compared with the 
calculations of \cite{WolMcK+95} .  Many properties of the model actually 
vary in concert including; the elemental and gas-phase C and O abundances
(there is no depletion); the columns of all other species heavier than 
Helium providing soft X-ray absorption; and the number of grains both large 
and small (the latter scales the heating rate of photoelectric ejection).  
\cite{Vla98} shows that the dust/metal ratio D/Z $\approx 0.6$ at 
higher redshift relative to conditions in the Milky Way (MW) so 
that D/G = D/Z $\times$ Z/G changes mostly because
of metallicity (Z/G), not D/Z.  In fact, such an overall scaling
has a surprisingly modest effect compared to some other changes
discussed next.  The warm phase may persist up to slightly higher
pressures and densities in gas of lower overall metallicity but
the existence of two-phase equilibrium is not threatened as long
as the balance between heating (small grains) and cooling (carbon)
is maintained in denser gas (as it is in this case).

At upper right, the gas-phase abundances of C and O are decreased
while all other parameters are held fixed and this actually has a
much more profound effect than scaling the grain (heating) and gas
(cooling) abundances in tandem.  Carbon provides some heating of the 
gas ($via$ its ionization, see \cite{SpiSco69}) but is more important 
as a source of electrons, influencing the grain heating and recombination 
rates, and as a coolant.  Clearly, when the ambient fluxes are
maintained while the main coolants 
(carbon and oxygen) and source of electrons are depleted, it is 
substantially more difficult to provide enough pressure to maintain 
gas in the cool phase.  This hearkens back to our discussion, slightly
earlier, of the lack of inclusion of depletion in the reference model 
for MW conditions.  In order to maintain two-phase stability over
the reference range of pressures
in the presence of gas-phase depletion, the incident optical/uv
ionizing radiation field must be decreased somewhat.  In fact this
might not be unreasonable locally for cool gas, which does not occur in 
arbitrarily small increments and therefore exists under conditions
of non-trivial extinction (especially at uv wavelengths).  At lower 
metallicity, this sort of compensation is less obvious and (consistent 
with some aspects of the observations) the pressure may in fact be 
somewhat higher in cool gas.

At lower left, we scale the ambient optical/uv radiation field which 
ionizes carbon (directly) and provides heating and electrons $via$ 
the photoelectric effect on small grains.  Regions whose ambient 
radiation field (ISRF) falls even a factor four below the average will 
be predominantly cool.  This is an inversion of the normal order of
things in diffuse gas and darker clouds in the nearby ISM occur at 
higher, not lower, thermal pressure.

Last, at lower right, we scale the soft X-ray flux.  Cool regions of 
higher density will probably never see the full extent of the ambient 
soft x-ray flux because even a small region of appreciable density
accumulates a high degree of X-ray absorption.  The heating
due to soft X-rays tends to increase on a per-event basis as the 
X-ray spectrum hardens 
somewhat after absorption, and tends to decrease as the electrons
cause more secondary ionization and less direct heating when
the ionization fraction is smaller (in neutral gas).

\subsection{Thermodynamics}

Fig. 2 shows, for the same set of parameter variations, the total
gas heating rate $\Gamma$ and the loss of energy (n$\Lambda$) due to 
cooling in the \158\ C II line (discussed below at some length), in
units of ergs s$^{-1}$ H$^{-1}$.  For the reference model, some 90\% 
of the heating is due to the photoelectron effect operating on small 
grains, see Fig. 3 of \cite{WolHol+95}. In warm gas, the burden of cooling 
is taken up by excitation of the \Lya\ lines under the
highly-idealized assumption that all photons escape; in fact, these photons
may travel substantial distances but by and large do not always 
escape a two-phase galactic layer before being absorbed by
dust, at Solar metallicity:  the situation may actually be very different 
in three-phase models, see \cite{Neu91} and \cite{Lis01}.  In cool gas, 
the cooling is due almost entirely to carbon and oxygen fine-structure
excitation, in the reference model. The curves of n$\Lambda$ in Fig. 2 
directly show the brightness (normalized; per H) of the C II* \158\
cooling transition ; in cool gas the O I$^*$ transition at \63\ 
makes up most of the cooling not provided by C II.

There is really a quite profound change in the local thermodynamics when the
metallicity or ionizing flux varies (Fig. 2, upper and lower left). In 
particular, while the energy input per H into cool and warm gas is nearly 
the same for the "Solar" metallicity and the standard ISRF or higher, it 
is much smaller in cooler gas when the metallicity is low.  In the Milky
Way, the carbon cooling rate can be used to infer the heating in both
cool and warm gas.  In systems of low metallicity such is not the case.

In general, the brightness of the C II $158\mu$ line scales with metallicity 
at all densities (Fig. 2, upper left) and varies nearly linearly in cool 
phase-stable gas since n(C II*)/n(C II) changes little with metallicity (see
Fig 5.).  It also scales with the strength of the ISRF in cool gas (Fig. 2, 
lower left), due to changes in n(C II*)/n(C II) (Fig. 5).  The C II \158\ 
brightness varies with depletion at lower densities (Fig. 2 upper right) and 
slightly, at low density, with variations in the soft X-ray flux.  
From the results at upper right in Fig. 2 we see that C II may not
be the dominant host of cooling in cooler gas, at quite high densities,
if the depletion (not the metallicity) is extreme;  oxygen becomes the 
preferred coolant due to the higher energy separation in its ground state
fine-structure levels.  There is no reason to believe that such conditions 
occur widely in diffuse gas, either locally or in \DLAS, although one 
source in Table 2 (PHL957) seems quite deficient in carbon, given its 
quoted metallicity.

\subsection{Carbon ionization}

Fig. 3 shows the ratio of the two lowest ionization states of carbon 
n(C II)/n(C I).  It differs by a factor of 40 - 100 between warm and 
cool neutral gas for the standard model, and is relatively insensitive 
to parameter changes, redshift, $etc.$  This makes it a sensitive 
indicator of the thermodynamic state of the gas, as discussed in Sect. 
3, but also causes confusion when mixtures of the two phases are observed 
along the same line of sight (see Sect 2.6).  Some aspects of the behaviour 
shown in Fig. 3 are counter-intuitive; for instance, a stronger ISRF maintains 
a high C II/C I ratio into denser gas, but the C II/C I ratio is actually 
$smaller$ in very tenuous gas when the radiation field is higher.  This 
arises in part because the equilibrium pressure increases with the 
strength of the radiation field and partly due to the level
of ionization in the gas (which is not solely determined by carbon).

If the C II lines are saturated or N(C II) is otherwise unknown, but N(C I) 
and N(H) are available, the thermodynamic state of the gas may also be 
inferred by computing N(C)/N(H) $\ga$ N(C II)/N(H) $\approx$ N(C II)/N(C I) 
$\times$ N(C I)/N(H).  Presumably, if the gas is actually cool, the 
inferred carbon abundance will be much too large (inconsistent with 
the metallicity) if the high n(C II)/n(C I) ratios typical of warm gas 
are assumed.

\subsection{C I fine-structure excitation}

The triplet fine-structure of the C I ground state is also accessible
to spectroscopy.  It is actually rather insensitive to the parameter
variations noted above (with the slight exception that C I*/C I 
increases modestly with increasing X-ray flux at low density) but 
responds very strongly to a change in the cosmic background level 
because the first excited state is only 23.2 K above ground.  Fig. 4 
(top) shows the ratio of populations in the lowest two levels as a 
function of \Tcmb.  Clearly, this ratio is a sensitive indicator of the 
ionization state at low redshift (it differs by a factor of about
thirty between warm and cool gas) but is rather insensitive for 
redshifts above 1.5 or so.  The C I*/C I ratio measures \Tcmb\ at high 
redshift in the context of our models (see \cite{RotMey92}).  Although 
perhaps within 
observational errors, it follows from the results in Fig. 4 that 
quoted column densities of neutral carbon for high-z systems should 
include a correction factor for the excited state population.

\subsection{Cooling rates and C II fine-structure excitation}

\begin{figure*}
\psfig{figure=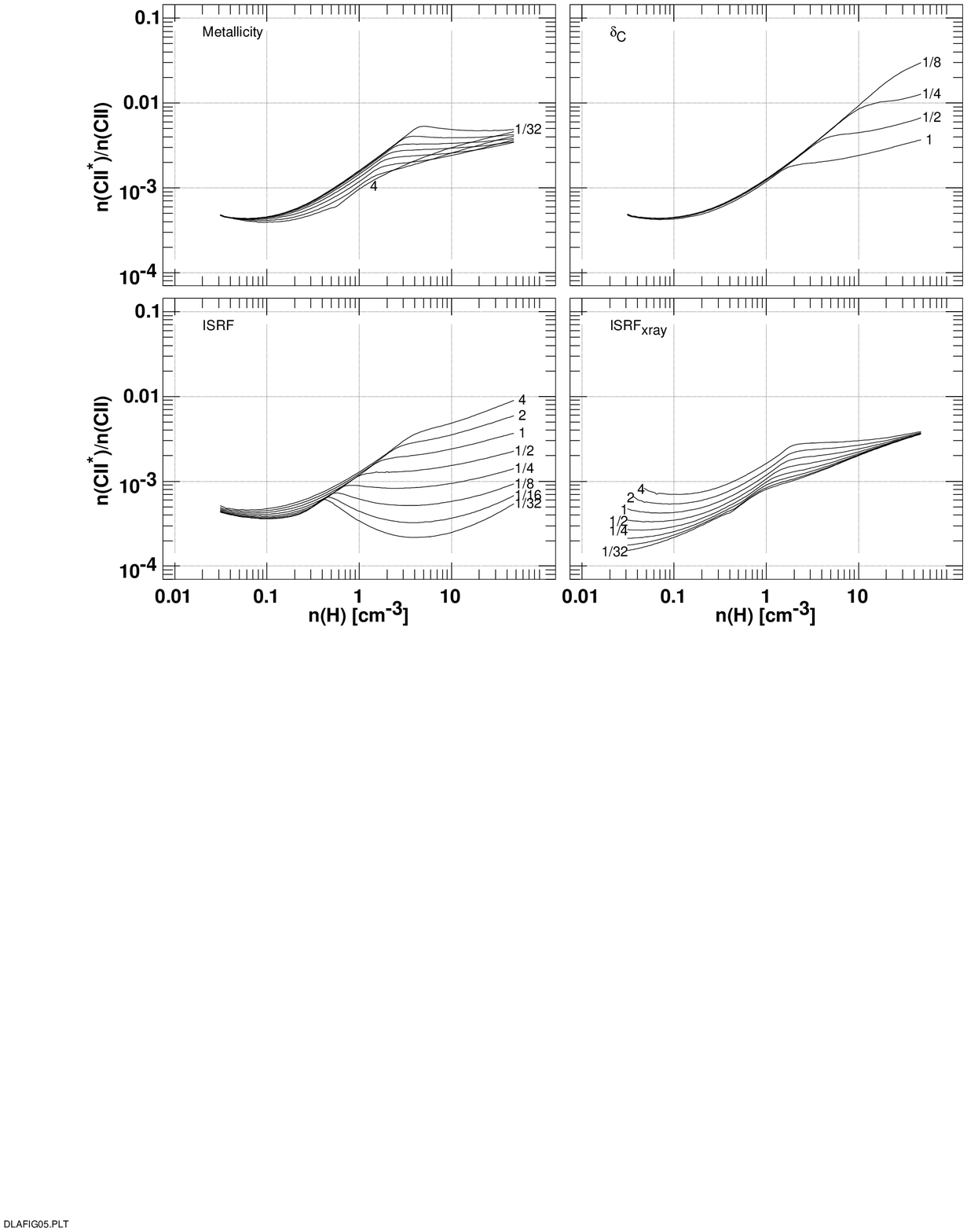,height=12.7cm}
\caption[]{As in Fig. 1, 2 and 3, but the quantity shown in each
panel here is the ratio of excited (C II*) to ground-state atoms in 
once-ionized carbon.}
\end{figure*}

\def\akj{\mbox{$_{\rm kj}$}}

The rate of energy loss due to a radiative transition between two
levels k and j having statistical weights g$_{\rm k}$ and g$_{\rm j}$,
level populations n$_{\rm k}$ and n$_{\rm j}$ (units of $\pccc$), 
energy separation E\akj\ and spontaneous emission rate A\akj\ ($\ps$), 
immersed in an isotropic radiation field characterized by \Tcmb\ is 
\citep{GolKwa74}

$$n \Lambda\akj = A\akj \beta\akj E\akj 
 (n_{\rm k} -(n_{\rm j} {\rm g}_{\rm k}/{\rm g}_{\rm j}
  -{\rm n}_{\rm k}){\rm F\akj(\Tcmb})) $$

where F\akj(x) $\equiv$ 1/(exp(E\akj/kx)-1).  $\beta$\akj\ is 
the probability of photon escape, typically varying as 
$\beta$\akj\ = (1-exp(-u$\tau$\akj))/(u$\tau$\akj) where u is a
geometrical factor of order unity and $\tau$\akj\ is the optical
depth of the transition.  The escape probabability is taken as
unity in these two-phase models, which is acceptable because the 
various fine-structure transitions do not achieve high optical depth
in typical diffuse clouds.  In calculation of the local
statistical equilibrium, these expresssions allow for a proper
evaluation of the fine-structure cooling rates for varying \Tcmb.

The 133.5 and 133.6 nm C II and C II* absorption lines have received
more attention because of the obvious possibility that C II is the
dominant ionization stage.  But discussions of the excitation of 
the C II* fine-structure levels often fail to recognize the role of 
carbon in determining the conditions under which it is observed; in many 
cases (see Fig. 2) C II will be the dominant coolant when it is the 
dominant ionization stage.  This means 
that the brightness of the C II \158\ line and the C II*/C II ratio are 
determined by thermodynamic equilibrium -- the energy input to the gas 
which occurs at a given density and temperature.  So, if the energy 
input into the gas does not change,  neither will the brightness of the 
\158\ line or the strength of the 133.5 nm C II* absorption line.  In Fig. 
2 at upper right, for a density of n(H) = 20 $\pccc$, it requires a factor 
8 drop in the the amount of carbon in the gas to weaken the \158\ line 
by about a factor 2, all other things being equal.  


Fig. 5 shows the parameter sensitivities of the C II*/C II ratio. Weak 
variation in the n(C II*)/n(C II) ratio in cool gas is shown in 
many ways in the panels of Fig. 5.  For the reference model, the ratio 
varies by only 50\% between n(H) = 2 $\pccc$ (\TK = 1000 K) and n(H) 40 
$\pccc$ (\TK = 75 K).  The variation with carbon depletion at upper
 right illustrates the effects of thermodynamic considerations. 
The ratio n(C II*)/n(C II) increases
as carbon is removed from the gas or n(C II) decreases, because the 
brightness of the \158\ line (and n(C II*)) must stay constant in order
to carry away a given amount of energy which is being input to the 
gas (see Fig. 2).  The behaviour at lower left, where the C II*/C II
ratio is a scaled representation of the cooling rate (also see
Fig. 2) shows how the brightness of the \158\ line, or the amount
of excited-state C II*, must change proportionally as the energy input 
to the gas changes.

In warm gas where its excitation is weak and C II is not the dominant
coolant, the population of the C II* level is susceptible to change 
with \Tcmb.  This is shown in the lower
two panels of Fig. 4.  The ability of the C II* population to
discriminate between phases may be lost at higher redshift.

\begin{table*}
\caption[]{Hydrogen and carbon at lower redshift}
{
\begin{tabular}{cccccccc}
\hline
Source & 3C286 & 0454+039 &
1756+237&1331+170&0013-004&1232+0815&PHL957 \\ 
z & 0.69 &0.86 &1.67&1.78&1.97&2.24&2.31 \\ 
\hline
$[$Zn/H$]$ & -1.22 & -1.1 &$(-0.8)^a$ &-1.27&-0.80&-1.20&-1.38\\ 
N(H I) & 1.8E21&4.2E20 &2.0E20&1.5E21&5.0E20&7.9E20&2.8E21 \\ 
N($\HH$) &&&&&6.9E19&6.3E16&$<$5.0E15 \\
N(C I) & $>$4E13& 4.4E13 &1.4E13&1.7E13&6.3E13&4.0E13 \\
N(C II) &&&&4.5E16&2.7E16&&7.8E15\\
N(C II*) &&&7.9E13&1.4E14&1.9E14&&7.6E13 \\
{N(C II)$_{\rm m}$}$^{ b}$ &3.8E16&1.2E16 & (1.1E16)&2.9E16&2.8E16&1.8E16&4.1E16 \\
N(C II)/N(C I)$^c$& ($<$952) & (293)& (803) &2647 &446 &445 &$>$1734 \\
N(C II*)/N(C II)&&& (0.0072)&0.0031&0.0070&&0.0097 \\

\hline
\end{tabular}}
\\
$^a$ assumed for illustrative purposes; other (entries) follow \\
$^b$ N(C II)$_{\rm m}$ is the amount of carbon if [C/Zn] is Solar \\
$^c$ values in parentheses use N(C II)$_{\rm m}$  \\
References: \\
3C286: \cite{BoiLeB+98,RotMey92} \\
0454+039: \cite{BoiLeB+98,SteBow+95} \\
1756+237: \cite{RotBau99} \\
1331+170: \cite{ChaFol+88,GeBec+97} \\
0013-004: \cite{GeBec+01} \\
1232+0815: \cite{BlaCha+87}; \cite{SriPet+00} quote N($\HH$) = 1.5E17$~\pcc$ \\
PHL957: \cite{BlaCha+87} \\
\end{table*}

In this context it is important to note that the $\Delta$E/k = 91 K 
energy separation of the C II fine-structure levels (to a lesser extent 
the 232 K separation in O I), is absolutely crucial to the similarity 
of two-phase equilibrium conditions out to moderate redshift, say
z $<$ 10.  If the gas locally were cooled by softer photons of 
energy $\Delta$E such that we could observe out to some redshift
where $\Delta$E/k\Tcmb $\la 4$, the distant gas would have to have
a different coolant.  It is apparent from the behaviour of the C I*/C I
ratio in Fig 4 that C I*, for example, could not be an important gas
coolant at z $>$ 3.   The same is true of the lower lines of CO (where 
$\Delta$E/k = 5.5 K, 11.0 K, 16.5 K, $etc.$).

\subsection{Observing admixtures of the two phases}

\def\RW{\mbox{R$_{\rm W}$}}
\def\RC{\mbox{R$_{\rm C}$}}
\def\Robs{\mbox{R$_{\rm OBS}$}}
\def\FC{\mbox{F$_{\rm C}$}}

When a property like the N(C II*)/N(C II) ratio -- call it R -- takes on 
values \RW\ and \RC\ in warm and cool gas, respectively, and a transparent
superposition of phases is observed to have a global value \Robs, 
\Robs\ is related to the proportion of gas in the two phases as

$$
  \FC = {{\RC}\over{\RW-\RC}}[{{\RW}\over{\Robs}}-1]
$$

where \FC\ is defined as the fraction of the gas in the cool phase.

\begin{figure}
\psfig{figure=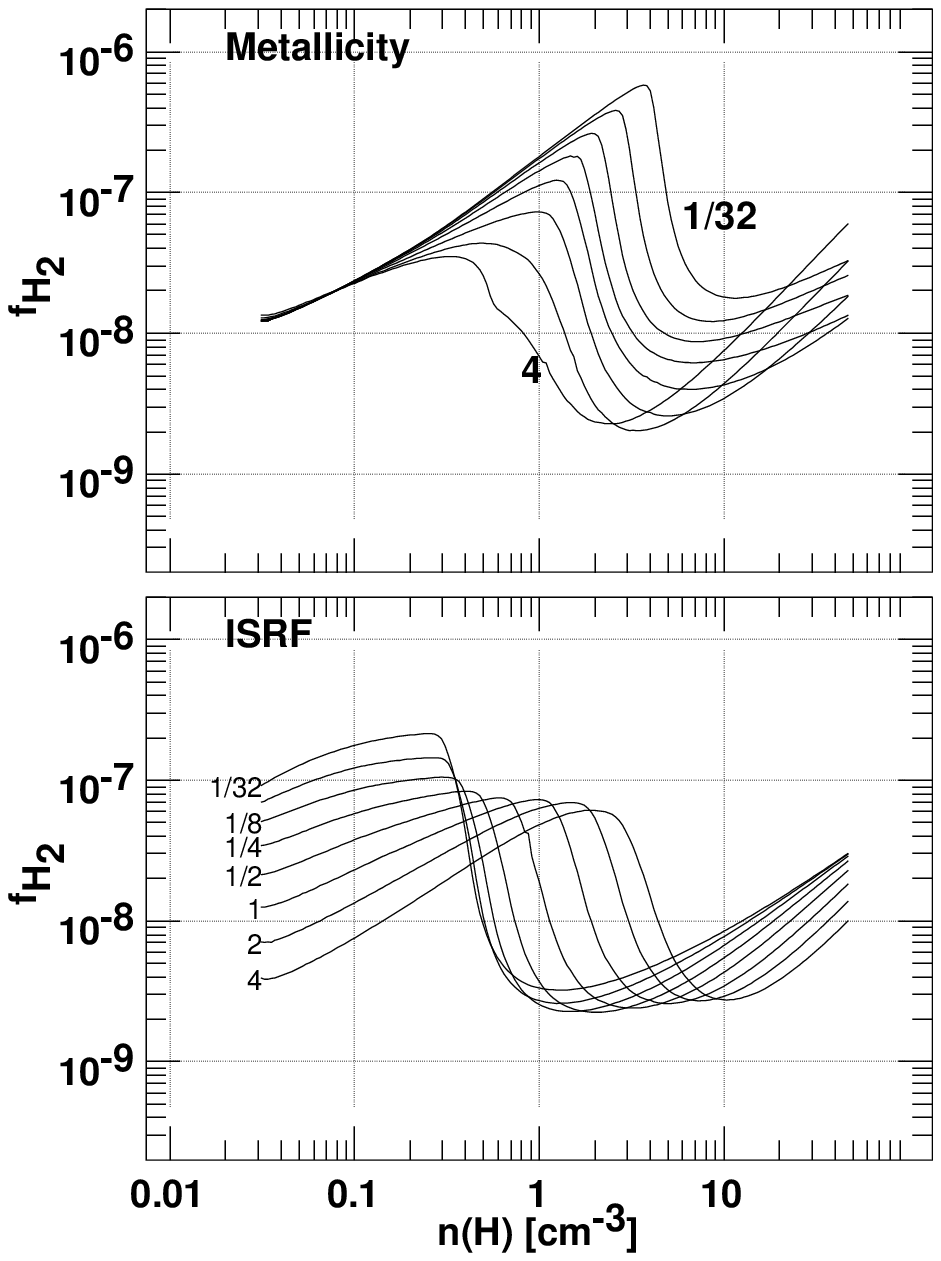,height=11.7cm}
\caption[]{Molecular fractions 2n($\HH$)/n(H) in free space resulting 
solely from slow gas-phase $\HH$-formation mechanisms, for the two-phase 
models and paramter variations shown at left in Fig. 1.  See Sect. 4 of 
the text.}
\end{figure}

In stable two-phase gas (Fig. 3 at upper left) \RW = 20000 and (roughly) 
\RC\ =  700/n(H) in gas of Solar metallicity (\RC\ varies rather less at 
low metallicity).  In this case, measuring a global average of, say, 
\Robs\ = 1000 implies only that there is some fraction \FC $\le 1$ of 
gas having \RC $\le 1000$ along the line of sight.

This ambiguity cannot generally be resolved with recourse to comparison of 
profiles, which are largely undifferentiated in the C II line, having been 
broadened due to some combination of; underlying kinematics in the host 
system; doppler-broadening of individual gas complexes; and, saturation.
In the Milky Way, C II lines are typically fit with much larger b-values 
(Doppler broadening parameters) even when a single cloud or gas complex 
dominates the line of sight to a nearby star ($cf.$ \cite{Mor75}).  
Because of this one cannot, for instance, try to compare the C II* with 
only that fraction of the C II which overlaps it in velocity.

\section{Warm or cool gas in \DLAS? -- observations of C I and C II}

In Tables 1 and 2 we have gathered information on the neutral and 
ionized carbon lines available in the literature.  For each source
we show the observed column densities, and the metallicity derived from 
zinc, which is believed to be (at most) only lightly depleted 
\citep{PetSmi+94,VlaBon+00}.  
From the metallicity we derive a quantity N(C II)$_{\rm m}$ which 
is meant to represent the maximum amount of carbon available in
the gas phase, i.e. it is the column density of carbon assuming
that carbon and zinc have the same metallicity and depletion.   
Comparison with the measured C II column densities shows that
carbon is $not$ lacking, with the notable exception of PHL957.
N(C II)$_{\rm m}$ is used as a surrogate where measured values 
of N(C II) are unavailable at low redshift in Table 1. 

The ratio N(C II)/N(C I) is much larger for the sources at high z
in Table 2, above 13,000-30,000, consistent only with an origin in 
overwhelmingly warm gas:  the fraction of cool gas must be of order 
a few percent at most. At lower redshift, two of the measured ratios, 
and that inferred for the source 1756+237, are below 1000, consistent 
with cool neutral gas in the stable region of the two-phase equilibrium, 
at thermal pressures P/k = 3,000 - 4000 $\pccc$ K, and densities 
n(H) = 10 - 70 $\pccc$.  The data for PHL957 are also consistent with 
stable cool gas at n(H) $\la 10~\pccc$, p/k $\la 2500~\pccc$ K, 
while 1331+170 is in the marginally unstable region at 
n(H) = $10~\pccc$, p/k = $2800~\pccc$ K.  For the systems at lower
redshift, there must be a very large proportion of cool gas, though
we cannot say exactly how much.

The N(C II*)/N(C II) ratios are noticeably smaller at high redshift in
Table 2, and would presumably be smaller still by a factor 2 or more,
were it not for the cosmic background radiation, which has
a noticeable effect (only) above z = 2.5 in warm gas (see Fig. 4).
The N(C II*)/N(C II) ratios observed at lower redshift, in presumably
cool gas, are 50-100\% higher than in the models.  This could be 
explained by an enhanced ISRF (Fig. 5, bottom left panel) or by some
depletion of carbon for PHL957, where the metallicity of carbon 
is (uniquely) very low compared to that of Zn, and N(C II*)/N(C II) is
rather large.

To summarize, comparison of the tabulated observational results with 
the calculated properties of carbon in two-phase media shows quite 
unambiguously that the observed gas has a substantial contribution from 
cool gas at lower redshift but is very largely warm at z $>$ 2.8.  Whether 
this shift from cool to warm gas is systematic or coincidental remains to be 
explored. \cite{NorSpa97} predicted that multi-phase equilibrium, cool 
gas and substantial quantities of $\HH$ would occur in protogalactic disks 
only when the metallicity had increased to 0.03 - 0.1 Solar, somewhere in the 
interval 1 $<$ z $<$ 2.

\section{The formation of molecular hydrogen}

\cite{LevDes+02} provide a summary of the molecular fractions
observed toward 14 sources (their Table 2) including most of those
discussed here.  The lowest molecular fraction is seen toward
0000-262 at z = 3.39, the highest toward 0013-004 at z = 1.97 
(see Tables 1-2 here).   The latter is the only direction for which the
molecular fraction exceeds 1/3000, and most are below $10^{-5}$.
In this Section we address the existence of \DLAS\ with large 
hydrogen column densities, substantial amounts of cool gas, but
very small fractions of molecular hydrogen. The models of 
$\HH$-formation presented here are very similar to those of 
\cite{LisLuc00}, who considered the formation of CO in local 
diffuse gas, given certain other conditions like the presence of 
\hcop, but there are a few differences which we now remark.

\subsection{$\HH$-formation in the gas phase}

The present calculations include X-ray heating and ionization,
because they are integral to the question of two-phase equilibrium.
X-rays have little effect on the large neutral gas columns which harbor 
appreciable molecular column densities nearby in the Milky Way but are
also included now because we are interested in understanding the minimum
molecular fractions which can be expected, and these are presumably
set by the slow gas-phase processes (not involving grains) which formed
the first stars, and which occur in low-density regions of (unshielded) 
free space.

\begin{table}
\caption[]{Hydrogen and Carbon at highest redshift}
{
\begin{tabular}{ccccc}
\hline
Source & 1337+113&0528-250&0347-382&0000-262\\
z &2.80&2.81&3.03&3.39\\
\hline $[$Zn/H$]$&-1.00&-0.91&-1.23&-2.07\\
N(H I) &8.0E20&2.2E21&2.52E20&2.6E21\\
N($\HH$)&$<$5.0E16&6.0E16&8.2E14&1.1E14\\
N(C I)&$<$1.6E13&$<$5.9E12&$<$4.0E11&\\
N(C II)&2.0E17&1.7E17&5.1E15&\\
N(C II*)&&3.6E14&1.9E13&\\
N(C II)$_{\rm m}$&2.8E17&9.6E16&3.1E15&7.9E15\\
C II/C I&$>$12658 &$>$28862 &$>$12700&\\
C II*/C II&&0.00211&0.00389&\\

\hline
\end{tabular}}
\\
References: \\
1337+113: \cite{LanWol+89} \\
0528-250: \cite{GeBec+97,SriPet98} \\
0347-382: \cite{LevDes+02} \\
0000-262: \cite{ProWol99,LevMol+00} \\
\end{table}

Gas phase $\HH$-formation occurs $via$ the exothermic reaction pathways
H +  e$^-$ $\rightarrow$ H$^-$ + $h\nu$, 
H$^-$ + H $\rightarrow$ $\HH$ + e$^-$ and
H$^+$ + H $\rightarrow$ $\HH^+$ + $h\nu$, 
$\HH^+$ + H $\rightarrow$ $\HH$ + H$^+$.
Many of the basic reactions are cited by \cite{PuyAle+93} and
most by \cite{HaiRee+96} (all can straightforwardly be located in the 
UMIST reaction database) but the discussion of early-universe conditions 
must be modified to include relevant values for the photodissociation of H$^-$
and $\HH$ and the cosmic-ray ionization of $\HH$.  For the latter we assumed
$\zeta = 2\times^{-17}\ps$ per H and 1.08 $\zeta$ per $\HH$), and for
the photodissociation rate of $\HH$ in free space we followed \cite{LeeHer+96}.
In order to formulate a treatment of the variation of the photodissociation 
of H$^-$ with extinction, we integrated the cross-sections of \cite{Wis79} 
over the local ISRF, finding an unshielded rate of 
$1.5\times 10^{-7}$ \ps, just over half of which arises at wavelengths 
beyond 800 nm; the photo-dissociation is, therefore, not strongly 
attenuated under the conditions discussed here and we elected to
ignore extinction in this regard (the rate quoted in the UMIST 
database is $2.4\times 10^{-7}$ \ps\ but we used the smaller value). 

Current values of the reaction rates for all important processes are 
given in the UMIST reaction database, whose values we employed unless
otherwise noted.  Species followed during modelling of the chemistry 
included H I, He$^+$, H$^+$ and e$^-$ -- all given by the calculations of
two-phase equilibrium --as well as H$^-$, $\HH$ and $\HH^+$.

\begin{figure}
\psfig{figure=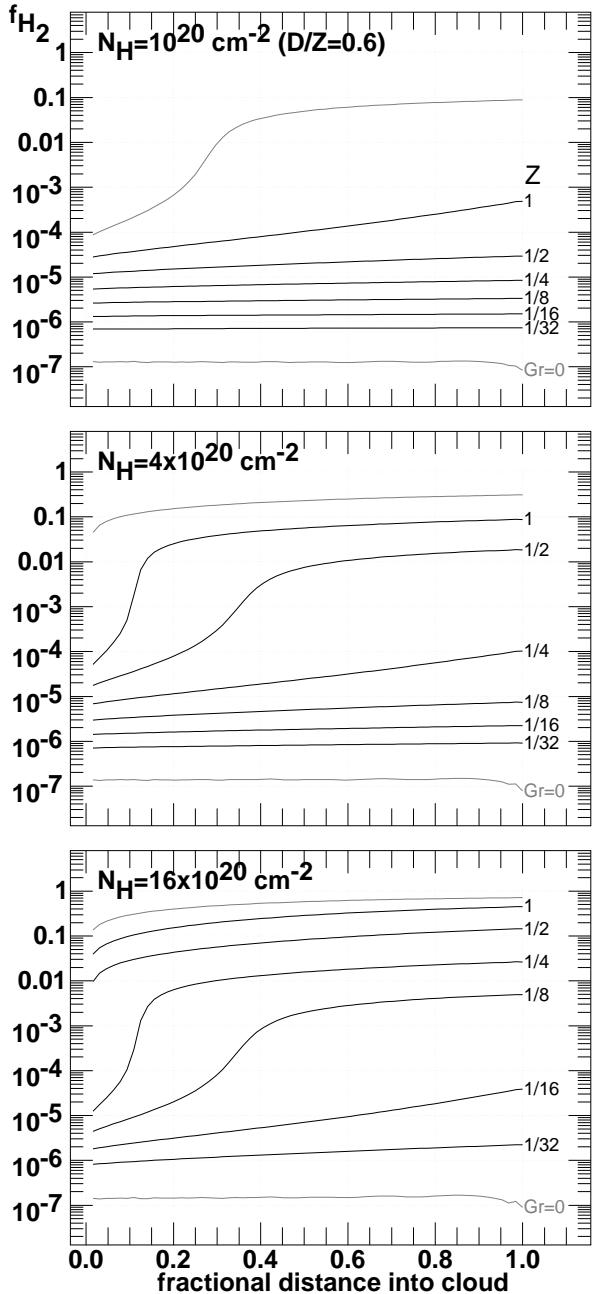,height=17cm}
\caption[]{Molecular fraction as a function of fractional distance 
into cool clouds of density n(H) $= 32~\pccc$ and various central 
column density N(H) as indicated.  The uppermost curve in each panel 
is for local conditions.  The labelled curves are for varying metallicity 
and a dust/metal ratio D/Z = 0.6 (60\% of the local value).  The bottom 
curve represents formation solely in the gas phase, in the limit of zero 
metallicity parameter; see Sect. 5 of the text.}
\end{figure}

Fig. 6 shows the free-space abundance of $\HH$ arising solely from 
gas-phase processes in the two-phase models under conditions of 
varying metallicity and ISRF (the two left-hand panels of Fig. 1).  
Unlike grain formation scenarios, which take advantage of high N(H) 
to boost the molecular fraction at high density, free-space gas-phase 
formation of $\HH$ does not seem to distinguish between warm and cool 
conditions.  The molecular fractions calculated in Fig. 6 correspond 
well with the smallest values in the local ISM, or, for that matter, 
in \DLAS.  Unfortunately, this minimum is not diagnostic of the
host gas conditions.   


\subsection{The $\HH$ abundance in cool clouds}

$\HH$ formation in cool neutral gas clouds is illustrated in Fig 7, 
which indeed shows why even \DLAS\ with appreciable cool gas still may
lack molecular hydrogen.  To create this diagram, we considered
(following \cite{LisLuc00}) a spherical clot of gas of constant
density, immersed in isotropic radiation fields (X-ray, cosmic-ray,
optical/uv, $etc.$).  This was computationally divided up into 128 
radial zones, in each of which we derived the temperature/ionization 
structure and $\HH$ abundance.  The latter requires iteration, because 
the maintenance of $\HH$ is a sharply non-linear process dependent on the 
$\HH$ column and extinction between any point and free-space 
\citep{LeeHer+96}.  We adopted a fairly straightforward relaxation 
method which converged with gratifying rapidity.

Calculation of the abundance of molecular hydrogen is typically made
feasible by employing a set of shielding factors which account in an
average way for the many very complicated effects of line-overlap and
radiation transport in the dissociation process.  We used the shielding 
factors of \citep{LeeHer+96} which were calculated for local gas.  The
justification for this is that the dominant effect requiring consideration 
here is the order of magnitude change in the number of grains at very
low metallicity, not the factor of two differences in individual grain
properties between local grains and those seen, for instance in the
Magellanic clouds.  The parametrization of \cite{Pei92} for the Milky
Way, LMC and SMC shows that, for a given amount of B-band extinction, 
the grain distribution provides successively somewhat more extinction at 
(say) Ly-$\alpha$ as the metallicity declines; the inference is that 
graphite grains disappear and silicates do not.  But this effect is 
dominated by the overall diminution of the extinction with lowered 
metallicity.

Figure 7 shows the radial variation of the fraction of H-nuclei in
molecular form over spherical gas clots of constant density n(H)
$= 32 ~\pccc$ for different column densities N(H) through the center
of the clot.  The mean line of sight averaged over the circular
face of such a body intersects it at an impact parameter of
2/3 of the radius (at a value 0.33 along the horizontal axis
in Fig. 7), where the column density is 3/4 of that through
the center. 

In each panel of the Figure there are 8 vertically-separated curves.
At top, shaded, is the result which would apply in the Milky Way, where
we have taken the dust/metal ratio as observed locally (the reference
model of the two-phase calculations) and depleted carbon and oxygen
in the gas phase by a factor of 2.4.  The bottom curve, also shaded, is
the result when the metallicity goes to zero and only the gas-phase
formation of $\HH$ is included; a modest amount of self-shielding
occurs and the molecular fraction is slightly higher than in free space 
(Fig. 6).  The intermediate curves assume a dust/gas ratio 0.6 relative 
to the reference model for the Milky Way following \cite{Vla98} (a small 
effect at higher N(H) but of real importance to the thinnest model) and 
no depletion of carbon and oxygen.  These curves are labelled by their 
varying metallicity as in the previous diagrams.

The cloud with N(H) = $4 \times 10^{20}~\pcc$ would be a compact (4 pc), 
cool (130 K) \cite{Spi78} ``standard'' cloud in the Solar neighborhood.  
It would also be very substantially molecular if found in the Solar 
vicinity, but a factor 4 decline in the metallicity suffices to 
reduce the molecular fraction by some four orders of magnitude.  Even
the model having a four times higher column density (compare with the
entries in Tables 1-2) cannot sustain an appreciable molecular fraction
when the metallicity is reduced by a factor 10, which is hardly extreme
for one of the \DLAS.

\subsection{The role of geometry}

The role of geometry can also be inferred from Fig. 7.  At any given
metallicity, a cloud with lower N(H) produces much less than one-fourth
as much $\HH$ as that illustrated in the next-lower panel.  This is
another reason why the molecular fraction may vary widely between two 
lines of sight with similar N(H), N(C II)/N(C I), and/or metallicity 
(for example).  Molecular hydrogen will readily populate a region when 
the circumstances are propitious, but can easily be prevented from
forming by the vagaries of local source structure.

\subsection{PKS0528-250}

This source (Table 2) has an overall molecular fraction $5.5\times10^{-5}$ 
despite the lack of evidence (in carbon) for any appreciable amounts of 
cool gas; earlier we asserted that (very roughly) no more than a few
percent of the gas could be cool.  \cite{CarLan+96} did not detect 
21cm H I absorption, placing a $2-\sigma$ upper limit 
N(H I) $\le 2 \times 10^{18}~\pcc\Tsp$. So, the molecule-bearing gas 
must be cool, occupying roughly 1\% of the total gas column for 
$\Tsp \la 100$ K.  In the context of our models the gas must also be
fairly dense, n(H) $\ga 100~\pccc$, occurring over only a very
small fraction of the path length (1 kpc or more ) occupied by the gas as a 
whole.

\section{Summary}

The thermodynamics of the ISM near the Sun are strongly influenced
by the disposition of carbon, in grains and in the gas.  Spectroscopy
of neutral and ionized carbon affords the opportunity to probe the 
processes which are most basic to the structure of the gaseous medium, 
nearby and in \DLAS.  For whatever reason, the very distant gas is easily 
understood in the same terms as that seen nearby.  Here, we have shown that 
one seemingly disparate aspect of high-z systems, their small fractions 
of molecular gas, can also be easily understood.  Even when cool gas is 
present, which must be the case for the systems discussed here at z $<$ 
2.3, abundances of $\HH$ are suppressed by many orders of magnitude at 
lower metallicity as a result of the sharply non-linear nature of the 
processes required to maintain substantial columns of $\HH$.  No wholesale 
reorganization of the gaseous medium need be hypothesized to account 
for low $\HH$ abundances.  Conversely, we showed that the slow gas-phase 
processes which formed $\HH$ in the early Universe provide for a minimum 
molecular fraction in the range $10^{-8}-10^{-7}$.

\begin{acknowledgements}

The National Radio Astronomy Observatory is operated by AUI, Inc. under 
a cooperative agreement with the US National Science Foundation.  The 
referee, Mark Wolfire, is thanked for helpful comments.

\end{acknowledgements}
 
\bibliographystyle{apj}
\bibliography{ms2497b1,ms2497b2}

\begin{thebibliography}{62}
\expandafter\ifx\csname natexlab\endcsname\relax\def\natexlab#1{#1}\fi

\bibitem[{{Bakes} \& {Tielens}(1994)}]{BakTie94}
{Bakes}, E. L.~O. \& {Tielens}, A. G. G.~M. 1994, ApJ, 427, 822

\bibitem[{{Balucinska-Church} \& {McCammon}(1992)}]{BalMcC+92}
{Balucinska-Church}, M. \& {McCammon}, D. 1992, ApJ, 400, 699

\bibitem[{{Bennett} {et~al.}(1994){Bennett}, {Fixsen}, {Hinshaw}, {Mather},
  {Moseley}, {Wright}, {Eplee}, {Gales}, {Hewagama}, {Isaacman}, {Shafer}, \&
  {Turpie}}]{BenFix+94}
{Bennett}, C.~L., {Fixsen}, D.~J., {Hinshaw}, G., {Mather}, J.~C., {Moseley},
  S.~H., {Wright}, E.~L., {Eplee}, R.~E., J., {Gales}, J., {Hewagama}, T.,
  {Isaacman}, R.~B., {Shafer}, R.~A., \& {Turpie}, K. 1994, ApJ, 434, 587

\bibitem[{{Black} {et~al.}(1987){Black}, {Chaffee}, \& {Foltz}}]{BlaCha+87}
{Black}, J.~H., {Chaffee}, F.~H., \& {Foltz}, C.~B. 1987, ApJ, 317, 442

\bibitem[{{Boisse} {et~al.}(1998){Boisse}, {Le Brun}, {Bergeron}, \&
  {Deharveng}}]{BoiLeB+98}
{Boisse}, P., {Le Brun}, V., {Bergeron}, J., \& {Deharveng}, J. 1998, A\&A,
  333, 841

\bibitem[{{Carilli} {et~al.}(1996{\natexlab{a}}){Carilli}, {Lane}, {de Bruyn},
  {Braun}, \& {Miley}}]{CarLan+96a}
{Carilli}, C.~L., {Lane}, W., {de Bruyn}, A.~G., {Braun}, R., \& {Miley}, G.~K.
  1996{\natexlab{a}}, Astron. J., 111, 1830

\bibitem[{{Carilli} {et~al.}(1996{\natexlab{b}}){Carilli}, {Lane}, {de Bruyn},
  {Braun}, \& {Miley}}]{CarLan+96}
---. 1996{\natexlab{b}}, Astron. J., 111, 1830

\bibitem[{{Carilli} {et~al.}(1996{\natexlab{c}}){Carilli}, {Lane}, {de Bruyn},
  {Braun}, \& {Miley}}]{CarLan+96b}
---. 1996{\natexlab{c}}, Astron. J., 112, 1317

\bibitem[{{Chaffee} {et~al.}(1988){Chaffee}, {Foltz}, \& {Black}}]{ChaFol+88}
{Chaffee}, F.~H., {Foltz}, C.~B., \& {Black}, J.~H. 1988, ApJ, 335, 584

\bibitem[{{Chengalur} \& {Kanekar}(2000)}]{CheKan00}
{Chengalur}, J.~N. \& {Kanekar}, N. 2000, Mon. Not. R. Astron. Soc., 318, 303

\bibitem[{{Fall} \& {Pei}(1993)}]{FalPei93}
{Fall}, S.~M. \& {Pei}, Y.~C. 1993, ApJ, 402, 479

\bibitem[{{Garmire} {et~al.}(1992){Garmire}, {Nousek}, {Apparao}, {Burrows},
  {Fink}, \& {Kraft}}]{GarNou+92}
{Garmire}, G.~P., {Nousek}, J.~A., {Apparao}, K. M.~V., {Burrows}, D.~N.,
  {Fink}, R.~L., \& {Kraft}, R.~P. 1992, ApJ, 399, 694

\bibitem[{{Ge} {et~al.}(1997){Ge}, {Bechtold}, \& {Black}}]{GeBec+97}
{Ge}, J., {Bechtold}, J., \& {Black}, J.~H. 1997, ApJ, 474, 67

\bibitem[{{Ge} {et~al.}(2001){Ge}, {Bechtold}, \& {Kulkarni}}]{GeBec+01}
{Ge}, J., {Bechtold}, J., \& {Kulkarni}, V.~P. 2001, ApJ, 547, L1

\bibitem[{{Goldreich} \& {Kwan}(1974)}]{GolKwa74}
{Goldreich}, P. \& {Kwan}, J. 1974, ApJ, 189, 441

\bibitem[{{Haehnelt} {et~al.}(1998){Haehnelt}, {Steinmetz}, \&
  {Rauch}}]{HaeSte+98}
{Haehnelt}, M.~G., {Steinmetz}, M., \& {Rauch}, M. 1998, ApJ, 495, 647

\bibitem[{{Haiman} {et~al.}(1996){Haiman}, {Rees}, \& {Loeb}}]{HaiRee+96}
{Haiman}, Z., {Rees}, M.~J., \& {Loeb}, A. 1996, ApJ, 467, 522

\bibitem[{{Hennebelle}(2000)}]{Hen00}
{Hennebelle}, P. 2000, Ph.D.~Thesis, Universit\'e Paris 7

\bibitem[{{Izotov} {et~al.}(2001){Izotov}, {Schaerer}, \&
  {Charbonnel}}]{IzoSch+01}
{Izotov}, Y.~I., {Schaerer}, D., \& {Charbonnel}, C. 2001, ApJ, 549, 878

\bibitem[{{Jenkins} {et~al.}(1983){Jenkins}, {Jura}, \&
  {Loewenstein}}]{JenJur+83}
{Jenkins}, E.~B., {Jura}, M., \& {Loewenstein}, M. 1983, ApJ, 270, 88

\bibitem[{{Kanekar} \& {Chengalur}(2001)}]{KanChe01}
{Kanekar}, N. \& {Chengalur}, J.~N. 2001, A\&A, 369, 42

\bibitem[{{Kritsuk} \& {Norman}(2002)}]{KriNor02}
{Kritsuk}, A.~G. \& {Norman}, M.~L. 2002, ApJ, 569, L127

\bibitem[{{Lane} {et~al.}(2000){Lane}, {Briggs}, \& {Smette}}]{LanBri+00}
{Lane}, W.~M., {Briggs}, F.~H., \& {Smette}, A. 2000, ApJ, 532, 146

\bibitem[{{Lanzetta} {et~al.}(1989){Lanzetta}, {Wolfe}, \&
  {Turnshek}}]{LanWol+89}
{Lanzetta}, K.~M., {Wolfe}, A.~M., \& {Turnshek}, D.~A. 1989, ApJ, 344, 277

\bibitem[{{Ledoux} {et~al.}(1998){Ledoux}, {Petitjean}, {Bergeron}, {Wampler},
  \& {Srianand}}]{LedPet+98}
{Ledoux}, C., {Petitjean}, P., {Bergeron}, J., {Wampler}, E.~J., \& {Srianand},
  R. 1998, A\&A, 337, 51

\bibitem[{{Lee} {et~al.}(1996){Lee}, {Herbst}, {Pineau Des Forets}, {Roueff},
  \& {Le Bourlot}}]{LeeHer+96}
{Lee}, H.~H., {Herbst}, E., {Pineau Des Forets}, G., {Roueff}, E., \& {Le
  Bourlot}, J. 1996, A\&A, 311, 690

\bibitem[{{Levshakov} {et~al.}(2002){Levshakov}, {Dessauges-Zavadsky},
  {D'Odorico}, \& {Molaro}}]{LevDes+02}
{Levshakov}, S.~A., {Dessauges-Zavadsky}, M., {D'Odorico}, S., \& {Molaro}, P.
  2002, ApJ, 565, 696

\bibitem[{{Levshakov} {et~al.}(2000){Levshakov}, {Molaro}, {Centuri{\' o}n},
  {D'Odorico}, {Bonifacio}, \& {Vladilo}}]{LevMol+00}
{Levshakov}, S.~A., {Molaro}, P., {Centuri{\' o}n}, M., {D'Odorico}, S.,
  {Bonifacio}, P., \& {Vladilo}, G. 2000, A\&A, 361, 803

\bibitem[{{Liszt}(2001)}]{Lis01}
{Liszt}, H. 2001, A\&A, 371, 698

\bibitem[{{Liszt} \& {Lucas}(1996)}]{LisLuc96}
{Liszt}, H.~S. \& {Lucas}, R. 1996, A\&A, 314, 917

\bibitem[{{Liszt} \& {Lucas}(2000)}]{LisLuc00}
---. 2000, A\&A, 355, 333

\bibitem[{{Lucas} \& {Liszt}(1996)}]{LucLis96}
{Lucas}, R. \& {Liszt}, H.~S. 1996, A\&A, 307, 237

\bibitem[{{McDonald} \& {Miralda-Escud{\' e}}(1999)}]{McDMir99}
{McDonald}, P. \& {Miralda-Escud{\' e}}, J. 1999, ApJ, 519, 486

\bibitem[{{Molaro} {et~al.}(2002){Molaro}, {Levshakov}, {Dessauges-Zavadsky},
  \& {D'Odorico}}]{MolLev+02}
{Molaro}, P., {Levshakov}, S.~A., {Dessauges-Zavadsky}, M., \& {D'Odorico}, S.
  2002, A\&A, 381, L64

\bibitem[{{Morton}(1975)}]{Mor75}
{Morton}, D.~C. 1975, ApJ, 197, 85

\bibitem[{{Neufeld}(1991)}]{Neu91}
{Neufeld}, D.~A. 1991, ApJ, 370, L85

\bibitem[{{Norman} \& {Spaans}(1997)}]{NorSpa97}
{Norman}, C.~A. \& {Spaans}, M. 1997, ApJ, 480, 145

\bibitem[{{Pei}(1992)}]{Pei92}
{Pei}, Y.~C. 1992, ApJ, 395, 130

\bibitem[{{Pei} {et~al.}(1991){Pei}, {Fall}, \& {Bechtold}}]{PeiFal+91}
{Pei}, Y.~C., {Fall}, S.~M., \& {Bechtold}, J. 1991, ApJ, 378, 6

\bibitem[{{Petitjean} {et~al.}(2000){Petitjean}, {Srianand}, \&
  {Ledoux}}]{PetSri+00}
{Petitjean}, P., {Srianand}, R., \& {Ledoux}, C. 2000, A\&A, 364, L26

\bibitem[{{Pettini} {et~al.}(1994){Pettini}, {Smith}, {Hunstead}, \&
  {King}}]{PetSmi+94}
{Pettini}, M., {Smith}, L.~J., {Hunstead}, R.~W., \& {King}, D.~L. 1994, ApJ,
  426, 79

\bibitem[{{Prochaska} \& {Wolfe}(1997)}]{WolPro97}
{Prochaska}, J.~X. \& {Wolfe}, A.~M. 1997, ApJ, 487, 73

\bibitem[{{Prochaska} \& {Wolfe}(1998)}]{WolPro98}
---. 1998, ApJ, 507, 113

\bibitem[{{Prochaska} \& {Wolfe}(1999)}]{ProWol99}
---. 1999, Astrophys. J., Suppl. Ser., 121, 369

\bibitem[{{Puy} {et~al.}(1993){Puy}, {Alecian}, {Le Bourlot}, {Leorat}, \&
  {Pineau Des Forets}}]{PuyAle+93}
{Puy}, D., {Alecian}, G., {Le Bourlot}, J., {Leorat}, J., \& {Pineau Des
  Forets}, G. 1993, A\&A, 267, 337

\bibitem[{{Roth} \& {Bauer}(1999)}]{RotBau99}
{Roth}, K.~C. \& {Bauer}, J.~M. 1999, ApJ, 515, L57

\bibitem[{{Roth} \& {Meyer}(1992)}]{RotMey92}
{Roth}, K.~C. \& {Meyer}, D.~M. 1992, BAAS, 24, 806

\bibitem[{{Spitzer}(1978)}]{Spi78}
{Spitzer}, L. 1978, Physical processes in the interstellar medium (New York
  Wiley-Interscience, 1978. 333 p.)

\bibitem[{{Spitzer} \& {Scott}(1969)}]{SpiSco69}
{Spitzer}, L.~J. \& {Scott}, E.~H. 1969, Astrophysics, 158, 161

\bibitem[{{Srianand} \& {Petitjean}(1998)}]{SriPet98}
{Srianand}, R. \& {Petitjean}, P. 1998, A\&A, 335, 33

\bibitem[{{Srianand} {et~al.}(2000){Srianand}, {Petitjean}, \&
  {Ledoux}}]{SriPet+00}
{Srianand}, R., {Petitjean}, P., \& {Ledoux}, C. 2000, Nature, 408, 931

\bibitem[{{Steidel} {et~al.}(1995){Steidel}, {Bowen}, {Blades}, \&
  {Dickenson}}]{SteBow+95}
{Steidel}, C.~C., {Bowen}, D.~V., {Blades}, J.~C., \& {Dickenson}, M. 1995,
  ApJ, 440, L45

\bibitem[{{Taylor} \& {Cordes}(1993)}]{TayCor93}
{Taylor}, J.~H. \& {Cordes}, J.~M. 1993, ApJ, 411, 674

\bibitem[{{Vladilo}(1998)}]{Vla98}
{Vladilo}, G. 1998, ApJ, 493, 583

\bibitem[{{Vladilo} {et~al.}(2000){Vladilo}, {Bonifacio}, {Centuri{\' o}n}, \&
  {Molaro}}]{VlaBon+00}
{Vladilo}, G., {Bonifacio}, P., {Centuri{\' o}n}, M., \& {Molaro}, P. 2000,
  ApJ, 543, 24

\bibitem[{{Vladilo} {et~al.}(2001){Vladilo}, {Centuri{\' o}n}, {Bonifacio}, \&
  {Howk}}]{VlaCen+01}
{Vladilo}, G., {Centuri{\' o}n}, M., {Bonifacio}, P., \& {Howk}, J.~C. 2001,
  ApJ, 557, 1007

\bibitem[{{Wishart}(1979)}]{Wis79}
{Wishart}, A.~W. 1979, Mon. Not. R. Astron. Soc., 187, 59P

\bibitem[{{Wolfe} \& {Davis}(1979)}]{WolDav79}
{Wolfe}, A.~M. \& {Davis}, M.~M. 1979, Astron. J., 84, 699

\bibitem[{{Wolfe} \& {Prochaska}(2000{\natexlab{a}})}]{WolPro00a}
{Wolfe}, A.~M. \& {Prochaska}, J.~X. 2000{\natexlab{a}}, ApJ, 545, 591

\bibitem[{{Wolfe} \& {Prochaska}(2000{\natexlab{b}})}]{WolPro00b}
---. 2000{\natexlab{b}}, ApJ, 545, 603

\bibitem[{{Wolfire} {et~al.}(1995{\natexlab{a}}){Wolfire}, {Hollenbach},
  {McKee}, {Tielens}, \& {Bakes}}]{WolHol+95}
{Wolfire}, M.~G., {Hollenbach}, D., {McKee}, C.~F., {Tielens}, A. G. G.~M., \&
  {Bakes}, E. L.~O. 1995{\natexlab{a}}, ApJ, 443, 152

\bibitem[{{Wolfire} {et~al.}(1995{\natexlab{b}}){Wolfire}, {McKee},
  {Hollenbach}, \& {Tielens}}]{WolMcK+95}
{Wolfire}, M.~G., {McKee}, C.~F., {Hollenbach}, D., \& {Tielens}, A.~G.~G.~M.
  1995{\natexlab{b}}, ApJ, 453, 673+

\end{thebibliography}

\end{document}